\documentclass[prb,twocolumn,aps,showpacs,10pt]{revtex4-1}

\usepackage{amsfonts}
\usepackage{amsmath}
\usepackage{graphicx}
\usepackage[latin1]{inputenc}
\usepackage{color}
\usepackage{bbold}

\newcommand{\sigmaV}{{{\mbox{\boldmath$\sigma$}}}}

\newcommand{\be}{\begin{equation}}
\newcommand{\ee}{\end{equation}}
\newcommand{\BE}{\begin{equation*}}
\newcommand{\EE}{\end{equation*}}
\newcommand{\bea}{\begin{eqnarray}}
\newcommand{\eea}{\end{eqnarray}}
\newcommand{\BEA}{\begin{eqnarray*}}
\newcommand{\EEA}{\end{eqnarray*}}

\newcommand{\w}{\omega}

\newcommand{\gb}{\bar \gamma}

\newcommand{\dd}{\dagger}

\newcommand{\ckad}{c_{k\alpha}^\dd}
\newcommand{\cka}{c_{k\alpha}}
\newcommand{\D}{\Delta}
\newcommand{\G}{\Gamma}
\newcommand{\Gt}{\widetilde{\G}}

\newcommand{\LL}{\Lambda}
\newcommand{\inn}{\infty}

\newcommand{\Gw}{\Gamma(\omega)}

\begin{document}
\title{Effect of a gap on the decoherence of a qubit}
\author{Juliana Restrepo$^{1}$, R. Chitra$^{1}$, S. Camalet$^{1}$ and \'Emilie Dupont$^{2}$.}
\affiliation{$^1$ Laboratoire de Physique Th\'eorique de la Mati\`ere Condens\'ee, UMR 7600, Universit\'e Pierre et Marie Curie, 4 place Jussieu, 75252 Paris Cedex 05, France. \\
 $^2$ Laboratoire de Physique Th\'eorique et Mod\'elisation, CNRS UMR 8089, Universit\'e de Cergy-Pontoise, F-95000 Cergy-Pontoise, France }
\begin{abstract}
We revisit the problem of the decoherence and relaxation of  a central spin 
coupled to a bath of conduction  electrons. We consider both metallic and semiconducting baths to
study the effect of a gap in the bath  density of states (DOS) on the time evolution of the density matrix of the
central spin. 
We use two weak coupling approximation schemes to study the decoherence. 
At low temperatures,  though the temperature dependence of  the decoherence rate  in the case of a metallic bath is the same irrespective of the details of the bath, the same is not true for  the semiconducting bath. 
We also calculate the relaxation and decoherence rates
as a function of external magnetic fields applied both on the central spin and the bath.  We find that in the presence of the gap,
 there exists a certain regime of fields, for which  surprisingly, the metallic bath has  lower rates of relaxation and decoherence than the semiconducting bath.

\end{abstract}
\date{\today} 
\maketitle
\section{Introduction}

 Solid state spin devices are  promising candidates for quantum computation \cite{Nielsen2000} and for quantum communication \cite{Bose2003}.  The basic building block  for quantum computation is the qubit, which is an effective two level system. Various realizations of qubits  include spins, Josephson junction qubits \cite{Makhlin2001} and others involving quantum dots \cite{Petta2004}. 
The efficiency of these systems as  qubits/quantum computing devices depends crucially on their coupling   to the numerous environmental degrees of freedom. Typically, environments are composed of phonons, electrons, nuclear spins or other noise inducing  objects and  lead to dissipation and a consequent loss of coherence of the qubit.  Understanding the dependence of this decoherence and relaxation on the physical parameters which define the environment is vital to fabricating good
qubits.   

The simplest models  used to study the effects of dissipation induced by the environment is
the Caldeira-Leggett   model \cite{Leggett1981},  and the  spin-boson\cite{Leggett1987} model where a  central spin  couples to  a bath of free bosons.   The past decade has however, seen a lot of studies  on various kinds of dissipative environments including interacting and non-interacting spin baths and electronic baths \cite{Camalet2007,Camalet2007a,Schiller2006,Stamp2004,Rossini2007,DasSarma2006} . In the limit of weak coupling between the central spin i.e., qubit  and the bath, 
intra-bath  interactions were sometimes seen to have a mitigating effect on the decoherence but this is not
 generically true \cite{Camalet2007}.   The  effect of fermionic
 environments  on single and multiple qubit systems has also been studied  in  \cite{Yamada2007,Schiller2006,Gao2008,Lutchyn2008}. 
 The  ensuing decoherence also depends on the nature of the coupling between the qubit and the bath fermions.
 In  general,   one expects Markovian decay of the coherence  of the qubit  at finite temperatures.   However, unlike the  case of bosonic baths, where the decoherence rate increases indefinitely with temperature,   the rate  saturates with temperature  for spin and fermionic baths. 
 This is however, not the case if there is long range order present in the bath or when the bath exhibits glassy features\cite{Camalet2007a,Winograd2009}.

 In this paper, we revisit the problem of decoherence induced by a simple fermionic environment
 weakly coupled to a qubit i.e., a central spin.
 We  consider a Kondo coupling  where  the central spin  interacts with  the local spin density of the
bath fermions at the central spin site and
  focus on the effect of different metallic and insulating densities of states, in particular,  that of  a direct gap in the density of states.   In the weak coupling limit, we  use the well known time-convolutionless (TCL) and Nakajima-Zwanzig (NZ) techniques \cite{Breuer}  to study decoherence.  Though the two methods are equivalent 
 for metallic density of states (DOS),   we see that these two approximations are  inequivalent for a
 semi-conducting DOS at low enough temperatures raising questions about the validity of the
 weak coupling approximations in certain limits \cite{Weiss}.   We find that in general the asymptotic 
 decoherence is qualitatively the same for a bath with a metallic DOS, though details of the DOS 
 are very relevant for determining the intermediate time behavior.  Furthermore, we also calculate the  relaxation and decoherence  rates,  $\gamma_1$ and $\gamma_2$  as a function of external magnetic fields
applied both on the environment and the central spin.Though the  metallic or semiconducting nature of the bath plays an important
 role at low temperatures,   the baths are qualitatively indistinguishable at higher temperatures.
  
%The paper is organized as follows: in Sec. \ref{sec:dynam}, we introduce the model and  the weak coupling formalism used to obtain  the decoherence and relaxation of the central spin.  We investigate the case of zero magnetic fields in Sec. \ref{zeromagnetic}. Our results for  the coherences of a metallic and a semiconducting bath at zero and finite temperatures are presented in Sec. \ref{sec:metal} and Sec. \ref{sec:semi}. In Sec. \ref{sec:spin}, we study the dependence of both the relaxation and decoherence rates on external magnetic fields applied on the qubit and bath respectively followed by a  brief discussion of our results  in Sec. \ref{sec:Diss}.
The paper is organized as follows: in Sec. \ref{sec:dynam}, we introduce the model and  the weak coupling formalism used to obtain  the decoherence and relaxation of the central spin.  We investigate the case of zero magnetic fields in Sec. \ref{zeromagnetic}. Our results for  the coherence of a qubit coupled to a metallic and a semiconducting bath at zero and finite temperatures are presented in Sec. \ref{sec:metal} and Sec. \ref{sec:semi}. In Sec. \ref{sectionnon}, we study the dependence of both the relaxation and decoherence rates on external magnetic fields applied on the qubit and bath respectively followed by a brief discussion of our results  in Sec. \ref{sec:Diss}.

\section{Weak coupling formalism}
\label{sec:dynam}
 We present the model Hamiltonian and the methodology used to study the reduced dynamics of 
 the central spin interacting with a bath.
   The Hamiltonian that describes a localized spin-$\frac1 2$ $\sigmaV_c$  coupled to a  bath of non-interacting electrons is given by:
\begin{align}
H = & H_S + H_B+  H_ {SB}\nonumber\\
H = & -h\sigma^z_c + \sum_{k\alpha}{\varepsilon_{k\alpha}\ckad\,\cka} -   \sigmaV_c.{\bf V}
\label{hamiltoniano}
\end{align}
where $H_S$ denotes the intrinsic 
Hamiltonian of the central spin   described by the Pauli matrices $\sigmaV_c$ subjected to a magnetic field $h$, $H_B$ denotes the bath of non-interacting electrons and  $H_{SB}$ describes the Kondo coupling between the central spin and the bath  with ${\bf V}=  \lambda \sum_{k,p,\alpha,\beta}c_{k\alpha}^{\dagger}\boldsymbol{\sigma}_{\alpha\beta}c_{p\beta}$  proportional to the spin density of the bath  electrons at the origin where the central spin is positioned.
The coupling constant $\lambda$  is smaller than all the other scales of the hamiltonian.
$\cka$,$\ckad$ are the fermionic annihilation and creation operators  with wave number $k$ and $\varepsilon_{k\alpha}=\varepsilon_k+\alpha H$ ($\alpha=\pm 1$) denotes the dispersion of the two spin species of the bath electrons subjected to an external magnetic field $H$.

\begin{figure}
\centering
\includegraphics[trim=0cm 0cm 0cm -0cm,width=8cm]{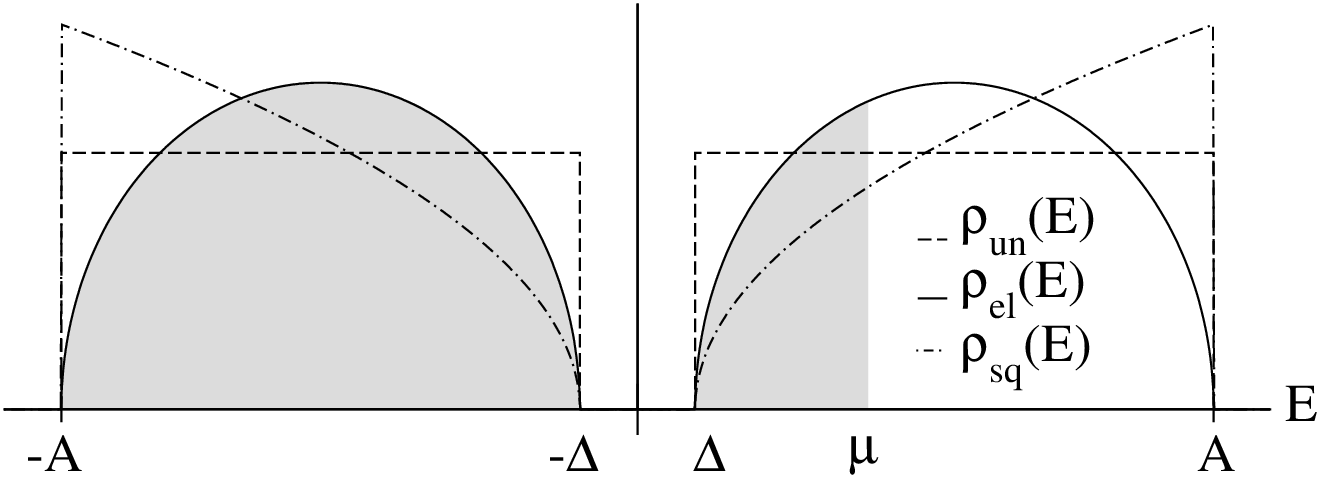}%{Figure.eps}%
\caption{Densities of states $\rho_{un}(E)$ (dashed lines), $\rho_{el}(E)$ (bold lines), $\rho_{sq}(E)$ (dot-dashed lines) as a function of $E$ for a metal $\mu>\D$.}
\label{densities}
\end{figure}
 
To  study the evolution (decoherence and relaxation) of the central spin, we assume  that at time $t=0$, the 
central spin is in a pure state $\lvert \psi \rangle = \alpha \lvert\uparrow\rangle + \beta \lvert\downarrow\rangle$  and that the  bath is  in thermal equilibrium at temperature $T$. This leads to 
a factorizable initial density matrix  $\Omega=\rho_s(0) \otimes \rho_B$ where 
 $\rho_s (0)=\lvert \psi  \rangle \langle \psi \rvert $  and 
 \begin{equation}
\rho_B=\frac{e^{- H_ B/T}}{Z} 
\end{equation}
where $Z=\mathrm{Tr}\left[  \exp \left(-\beta H_B \right)\right]  $ is the bath partition function. We use units $\hbar=k_B=1$ throughout this paper.
The time evolution of the reduced density matrix of the central spin is given by 
\begin{equation}\label{densm}
 \rho_s (t) = \mathrm{Tr}_B \left[ e^{-iHt} \Omega e^ {iHt} \right]
\end{equation} 
where $\mathrm{Tr}_B$  is the partial trace over  the bath degrees of freedom.
 We   rewrite the  density matrix (\ref{densm}) as a Laplace transform
\begin{eqnarray}
\rho_s(t) &\equiv &\frac{i}{2\pi} \int_{R+i\eta} dz  e^{-izt}  \rho_s(z) \nonumber  \\
&=&
  \frac{i}{2\pi} \int_{R+i\eta} dz  e^{-izt}  
\mathrm{Tr}_B \left[  \left( z-{\cal L} \right)^{-1} \rho(0) \otimes\rho_B \right] \label{rhot} 
\end{eqnarray}  
where $\eta$ is a real positive number and ${\cal L}$ is the Liouville operator corresponding to the total  Hamiltonian $H$, i.e., ${\cal L} A = \left[H, A\right]$ for any operator $A$. The density matrix can be decomposed in the basis of Pauli spin operators as follows:
\begin{eqnarray}
\rho_s(z) &=& \frac{1}{2} \sum_{\alpha=0,x,y,z} M_{\alpha} (z) \sigma_c ^{\alpha}
\end{eqnarray} 
The decoherence and relaxation are given by the quantities $M_\alpha$ which are  in general rather
difficult to calculate. However, for weak   central spin-bath coupling i.e., $\lambda \to 0$, we can use perturbation theory in conjunction with the projection operator technique  to calculate the components $M_\alpha$ \cite{Dattagupta1989}. These lead to  equations for $M_\alpha$  which go beyond   the standard Markovian master equation. To second order in $\lambda$, the $M_\alpha(z)$ satisfy the following matrix equation
\begin{equation}
 z M_{\beta} (z) - \sum_{\alpha} h_{\beta\alpha} M_{\alpha} (z) -  \sum_{\alpha} \Sigma_{\beta\alpha} (z) M_{\alpha} (z) = \langle \sigma_c ^{\beta} \rangle_{0}
\label{eqM}
\end{equation} 
where $\Sigma_{\alpha\beta} $  are the self-energies and only $h_{xy}=h_{yx}^{\ast}=2ih$ are nonzero.
In the absence of  external magnetic fields ($h=H=0$), since  we have  a SU(2) invariant bath (i.e.,
 $\varepsilon_{k\alpha}=\varepsilon_{k-\alpha}$),  the self energy matrix    simplifies with the only non-zero entries being the diagonal terms   $\Sigma_{x x} =\Sigma_{yy} =\Sigma_{zz} \equiv \Sigma$. These diagonal self-energies are related to the  bath correlation function
  \begin{eqnarray}\label{sigcor}
\Sigma (z) = -8 i  \int_{0} ^{\infty} dt e^{izt} 
\left[\Re e{ \langle \tilde{V}_{x}(t) \tilde{V}_{x} \rangle} \right] \; .
\end{eqnarray} 
where  for any operator $A$, $\tilde A$ is defined as $\tilde{A}(t)=e^{iH_B t}Ae^{-iH_B t} - \langle A \rangle$ and $\langle \ldots \rangle$ denotes the thermal expectation value, i.e.,$\langle A \rangle=\mathrm{Tr}(A \rho_B)$ for any operator $A$.  Consequently,   $M_x=M_y=M_z\equiv M $  with $M$ being given by
\begin{equation}
M(z)= \frac1 {z- \Sigma(z)}
\end{equation}
\noindent
The exact expressions of $h_{\alpha\beta}$ and $\Sigma_{\alpha \beta}$ for the general case are detailed in the  appendix.\\\\
To obtain $M(t)$ and hence $\rho_s(t)$,  we first analytically continue $\Sigma$ to real frequencies $\omega$
 \begin{equation}
 \lim_{\eta \rightarrow 0^{+} } \Sigma (\omega + i \eta )= \Lambda (\omega) -i \Gamma (\omega)
 \end{equation} 
 where   $\Lambda$  and $\Gamma$ are related by the  Kramers-Kronig relation.
 \begin{equation}
 \Lambda=-\frac{1}{\pi}P\int{\frac{\G(\w')}{\w'-\w}d\w}
\label{lambdita}
\end{equation}
Note that $\Gamma$ is even in $\omega$ whereas $\Lambda$ is odd.
For the case of the electronic bath studied here,  calculating the correlation in Eq.(\ref{sigcor}),  we find that  
$\Gamma$  is given by
 \begin{eqnarray}
\Gamma (\omega)=8\pi\lambda^2\left[S(\omega)+S(-\omega)\right]
\label{gamma}
\end{eqnarray} 
where $S(\omega)$ is the dynamical spin structure factor for the bath and given by 
\begin{equation}\label{sw}
S( \omega)=\int{d E\rho(E)\rho(E+\omega)n(E)\left[ 1-n(E+\omega)\right] }
\end{equation}
In Eq.(\ref{sw}), $\rho(E)$  is the   density  of states (DOS) for the bath  and  $n(E)$ is the Fermi occupation number.
 
 The inverse Laplace transform of  $M(z)$ directly  gives the Nakajima-Zwanzig (NZ)  approximation for the coherence
 \begin{eqnarray}
\ M_{NZ}(t) &=& \dfrac{1}{ \pi} \int_{0}^{\infty} d\omega \cos (\omega t) \tilde{\Gamma} (\omega) 
 \label{eq:M_NZ} 
\end{eqnarray} 
where
\begin{equation}
 \tilde{\Gamma} (\omega) = \Gamma  (\omega)/ \left[ (\w - \Lambda (\omega))^2 + \Gamma (\omega) ^2 \right]
\label{gtilde}
\end{equation}
\noindent
  On the other hand, if the self energy $\Sigma(z)$ is analytic in the lower half plane\cite{Camalet2007a}, we obtain a simplified form for the
coherence which is  the well known time convolutionless approximation (TCL)  
 \begin{eqnarray}
\ln  M_{TCL} (t) &\simeq& -\dfrac{2}{\pi} \int_{-\infty}^{\infty} d\omega \dfrac{\sin (\omega t /2) ^2}{\omega^2}  \Gamma (\omega) \label{eq:M_TCL}  \end{eqnarray} 
\noindent
Note that the NZ  approximation corresponds to a master equation which is non-local in time
i.e., it has a memory kernel,  whereas   TCL approximation corresponds to a purely local in time master equation  \cite{Breuer2004}.  Though the TCL is local in time,  the time-integration over the  memory kernel is  the characteristic feature which can capture  the non-Markovian nature of the dynamics.
If  $\Sigma(z)$ is  analytic in the lower half plane, both techniques lead to the same asymptotic behavior for the coherence but they  might yield
different results in cases where $\Sigma(z)$ is not  analytic. As we shall show later,
the two methods converge for the case of metallic DOS wherein $\Sigma$ is analytic in the lower half plane,
but they predict different results  for the simple case of a semiconducting DOS, because
of the presence of weak non-analyticities on the real axis.

In the presence of external magnetic fields,  the self-energy matrix $\Sigma_{\beta\alpha}$ has a far more complicated structure  than the one described above and requires a full matrix inversion to obtain the coherences $M_\alpha$.
However, the two timescales  $T_1$ (rate $\gamma_1$) and $T_2$ (rate $\gamma_2$) which determine the asymptotic relaxation and decoherence  i.e., $\langle \uparrow \lvert \rho_S (t) \lvert \uparrow \rangle \sim \exp{[-t/T_1]} $ and $\langle \uparrow \lvert \rho_S (t) \lvert \downarrow \rangle \sim \exp{[-t/T_2]} $   are given 
   by the bath correlation functions as\cite{Abragam1994}:
 \begin{eqnarray}
\gamma_1=\dfrac{1}{T_1}&=& \int dt e^{2iht} \left[\langle \tilde{V}_{+}(t) \tilde{V}_{-}\rangle + \langle \tilde{V}_{-} \tilde{V}_{+}(t) \rangle  \right] 
\label{eq:T1} \\
\gamma_2=\dfrac{1}{T_2}&=&\dfrac{1}{2T_1} + 2  \int dt \Re e \langle \tilde{V}_{z}(t) \tilde{V}_{z}\rangle
\label{eq:T2}
\end{eqnarray}
with $\tilde{V}_{\pm}=\tilde{V}_{x} \pm i \tilde{V}_{y}$.In the rest of the paper, we use  the dimensionless rates  ${\bar{\gamma}}_{1,2}=\gamma_{1,2}/8\pi\lambda^2$. Note that  for $h=H=0$,   we have $\gb_1=\gb_2=\gb$. 

\section{Zero magnetic field}
\label{zeromagnetic}
We now  use the weak coupling formalism  presented in the previous section to calculate the coherence
$M(t)$ for fermionic baths having both metallic and insulating density of states.  
In order to study the effect of the details of the DOS on the asymptotic coherence, we consider three densities of states: uniform, square root and elliptical,  each of which has a gap $\D$ and a cut-off $A$:
\begin{eqnarray}
 \rho_{un}(E) &=&\frac{1}{2(A-\D)}\Theta(|E|-\D)\Theta(A-|E|) \nonumber \\
 \rho_{sq}(E) &=& N\sqrt{|E|-\D}\Theta(|E|-\D)\Theta(A-|E|)\nonumber \\
 \rho_{el}(E)&=&N'\left(\frac{(A-\D)^2}{4} - \left(|E|-\frac{(A+\D)}{2}\right)^2\right)^{\frac12}
\end{eqnarray}
where the normalization  constants $N=\frac{3}{4}(A-\D)^{-\frac32}$
and  $N'= \frac{16}{3\pi}N(A-\D)^{-\frac12}$ have been  chosen  so that all densities of states have the same normalization. 
Depending on the chemical potential $\mu$, one obtains a metallic bath for $\mu >\D$ and a semiconducting  bath for $ \mu <\D$.
Though real semiconductors have more complicated DOS, we nonetheless expect the simple
DOS used in this paper to  capture the essential features of the asymptotic relaxation and 
decoherence induced by such baths.
  The gaps in  semiconductors are typically of the order of $1$ eV: $1.12$ eV for Silicon and $0.66$ eV for Germanium\cite{Kittel} and the cut-off $A$ is  of the order of $10$ eV.

\begin{figure}
\centering
\includegraphics[width=8cm]{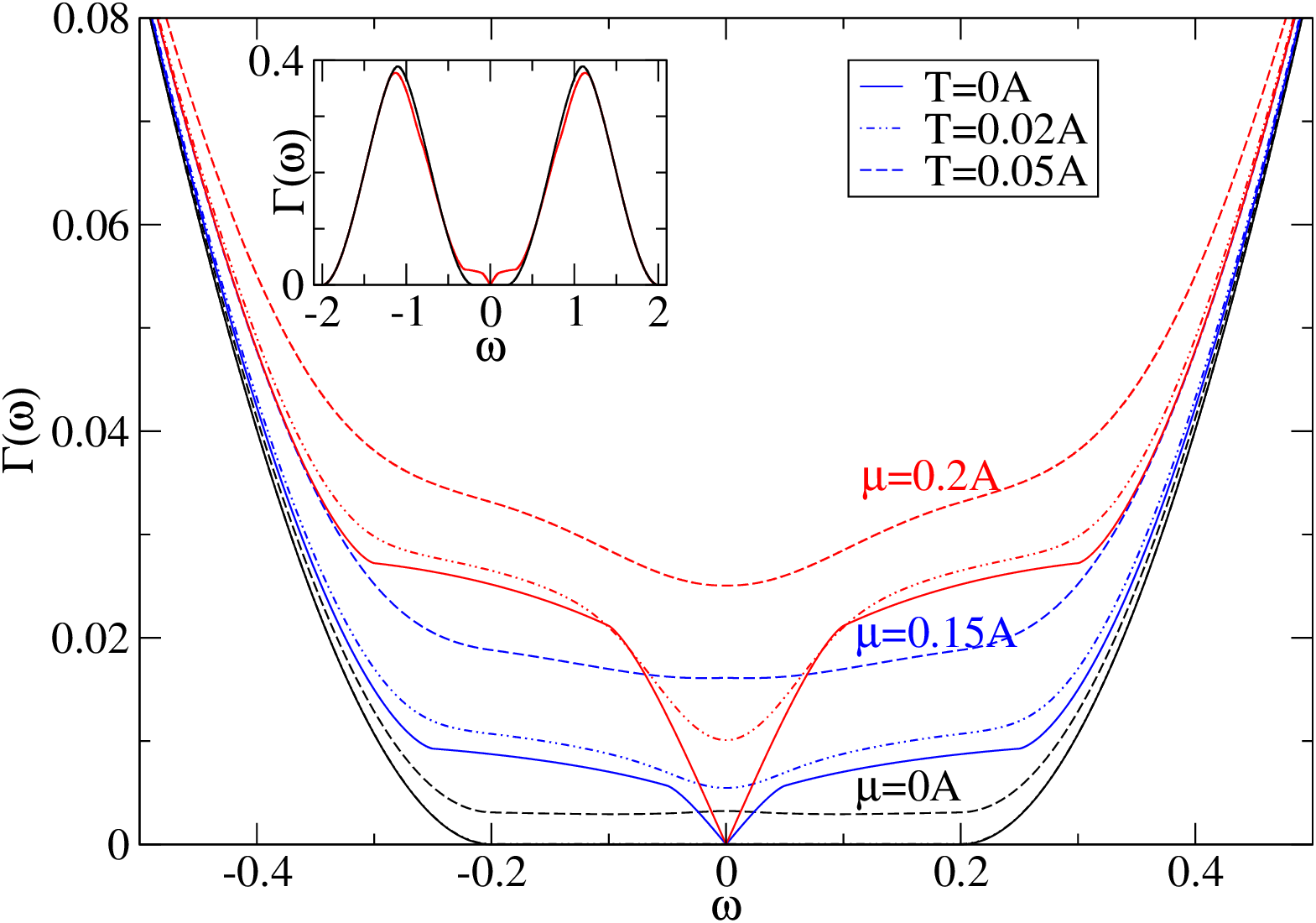}
%OLD\caption{(color online) Low frequency behavior of $\Gw/8\pi\lambda^2N^2$ as a function of $\w$ (units of $A$) for different values of $\mu$ and $T$ for the elliptical density with $\D=0.1A$. In the Figure, the temperatures $T=0$, $T=0.02A$ are indistinguishable for $\mu=0A$. The inset shows the full $\Gw/8\pi\lambda^2N^2$ for a metal at  $\mu=0.2A$ and a semiconductor $\mu=0A$ at zero temperature.}
\caption{(color online) Low frequency behavior of $\Gw/8\pi\lambda^2N^2$ as a function of $\w$ (units of $A$) for different values of the chemical potential: $\mu=0$ (black), $\mu=0.15A$ (blue), $\mu=0.2A$ (red) and temperature: $T=0$ (bold lines), $T=0.02A$ (dot-dashed lines), $T=0.05A$ (dashed lines) for the elliptical density with $\D=0.1A$. In the Figure, the temperatures $T=0$, $T=0.02A$ are indistinguishable for $\mu=0A$. The inset shows the full $\Gw/8\pi\lambda^2N^2$ for a metal at $\mu=0.2A$ (red) and a semiconductor $\mu=0$ (black) at zero temperature.}
\label{g_figure}
\end{figure}

\subsection{Metallic bath}
\label{sec:metal}

Here we assume that the  chemical potential $\mu > \D$.
At zero temperature and  for small frequencies $\omega$,  using Eqs. (\ref{gamma}) and (\ref{sw}),
we obtain the generic result
$\Gw=2K|\w|$ reminiscent of the Ohmic behavior seen in the standard spin-boson problem \cite{Leggett1987}.  The coefficient $K$ depends on the DOS and we find that
 $K_{un}=\pi\lambda^2/(A-\D)^2$, $K_{sq}=4\pi\lambda^2N^2(\mu-\D)$ and  $K_{el}=4\pi\lambda^2N^2\left(\frac{16}{3\pi}\right)^2(A-\D)^{-1}(\mu-\D)(A-\mu)$.  The fact that $K_{un}$ is independent of $\mu$ is an artifact of the uniform DOS.  This Ohmic  bath leads to the expected power law decay of the decoherence
$M(t)\simeq t^{-4K/\pi}$ where for the different DOS, the appropriate value of $K$ should be
substituted.  The asymptotic  decoherence induced by a metallic bath  is  qualitatively similar to that induced by  a bosonic bath at $T=0$.  This  correspondence with  the spin-boson problem does not hold  true  at finite temperatures or in the presence of  magnetic fields as we will show later. The full behavior of $\Gamma(\omega)$ for the case of the elliptical DOS is plotted in 
 Fig \ref{g_figure} for various temperatures and chemical potentials.  \\\\

\begin{figure}[ht]
\centering
\includegraphics[width=8cm]{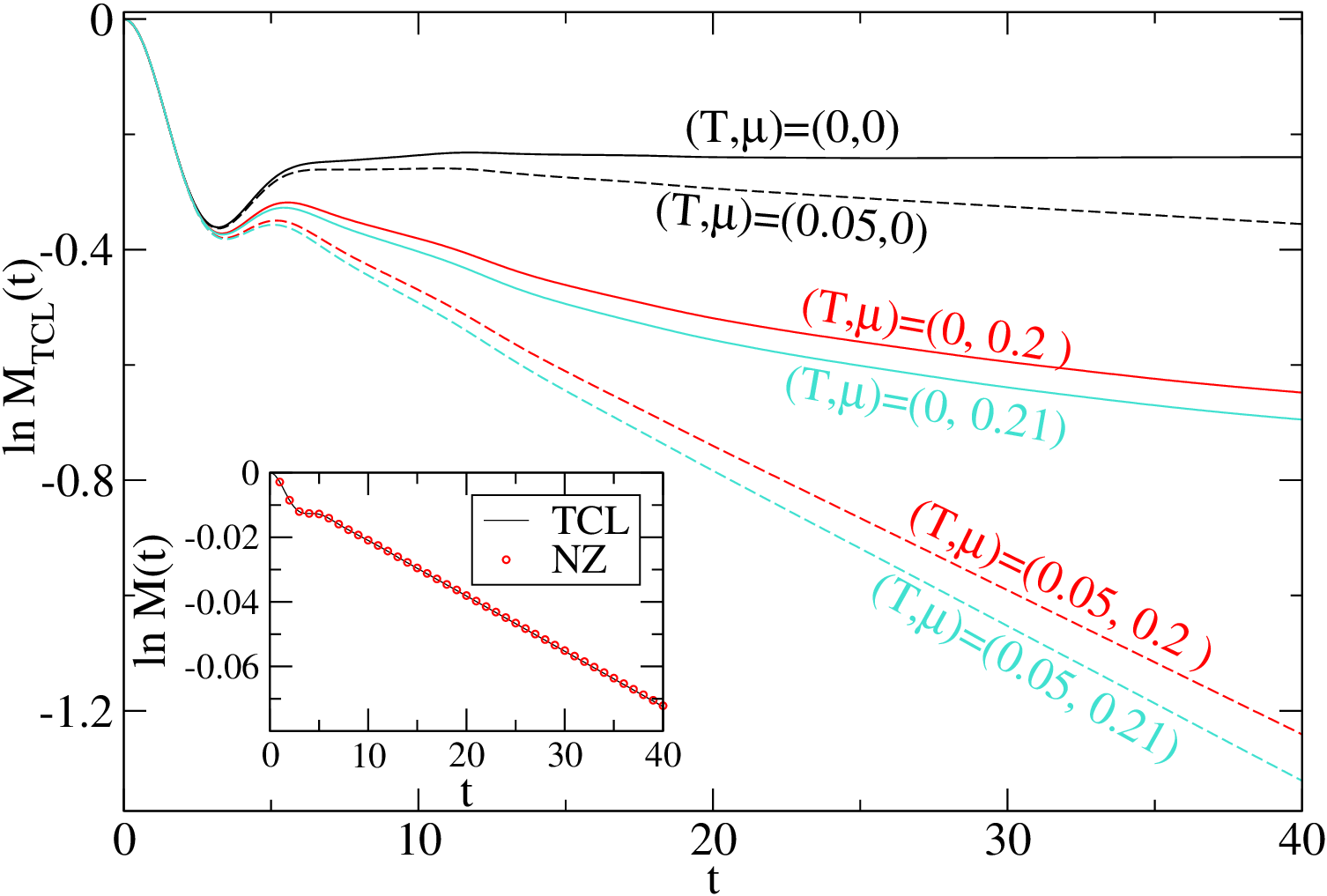}
%OLD\caption{(color online) $\ln M_{TLC}/8\pi\lambda^2N^2$ as a function of $t$ for a $\mu=0, 0.2, 0.21$ and $T=0, 0.05$ for the elliptical DOS with $\D=0.1$. In the inset we plot $\ln M/N^2$ for $NZ$ and $TCL$ approximations for $(T,\mu)=(0.1,0.2)$ and $\lambda=0.05$. Time, gap and chemical potential are in units of $A$.}
\caption{(color online) $\ln M_{TLC}/8\pi\lambda^2N^2$ as a function of $t$ for a $\mu=0$ (black), $0.2$ (red),  $0.21$ (turquoise) and $T=0$ (bold-lines), $0.05$ (dashed-lines) for the elliptical DOS with $\D=0.1$. In the inset, we plot $\ln M/N^2$ for $NZ$ and $TCL$ approximations for $(T,\mu)=(0.1,0.2)$ and $\lambda=0.05$. Time, gap, chemical potential and coupling strength are in units of $A$.}
\label{lnm}
\end{figure}

\begin{figure}[ht]
\centering
\includegraphics[trim=0cm 0cm -2cm -2cm,width=8cm]{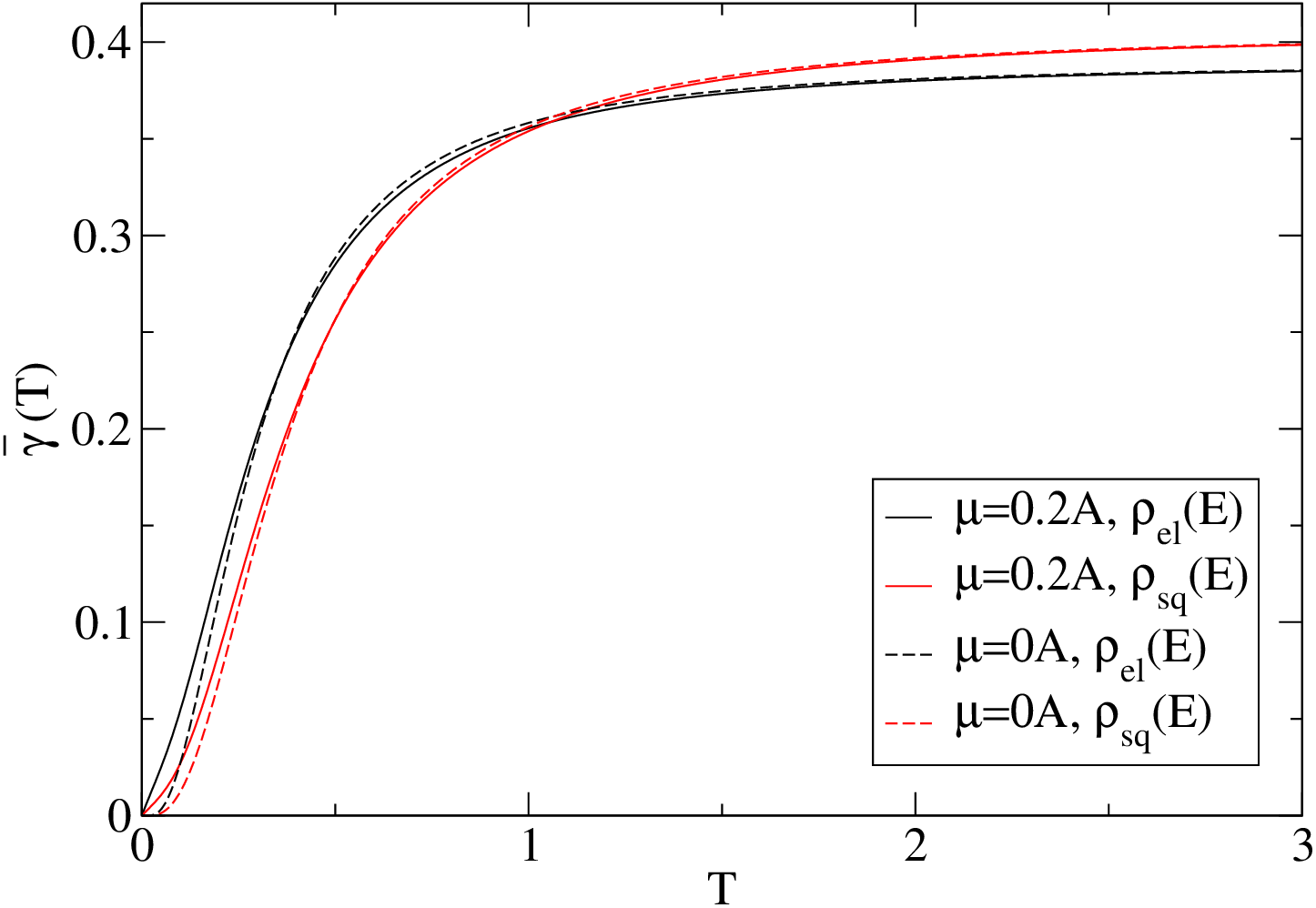}
\caption{(color online) Dimensionless rate $\bar{\gamma}/N^2$ as a function of temperature in units of $A$ for a semiconductor $\mu=0$ (dashed lines) and a metal $\mu=0.2A$ (bold lines) for the elliptical  (black) and square root (red) DOS with $\D=0.1A$. }
\label{figura4_v7}
\end{figure}

\begin{figure}
\centering
\includegraphics[width=8cm]{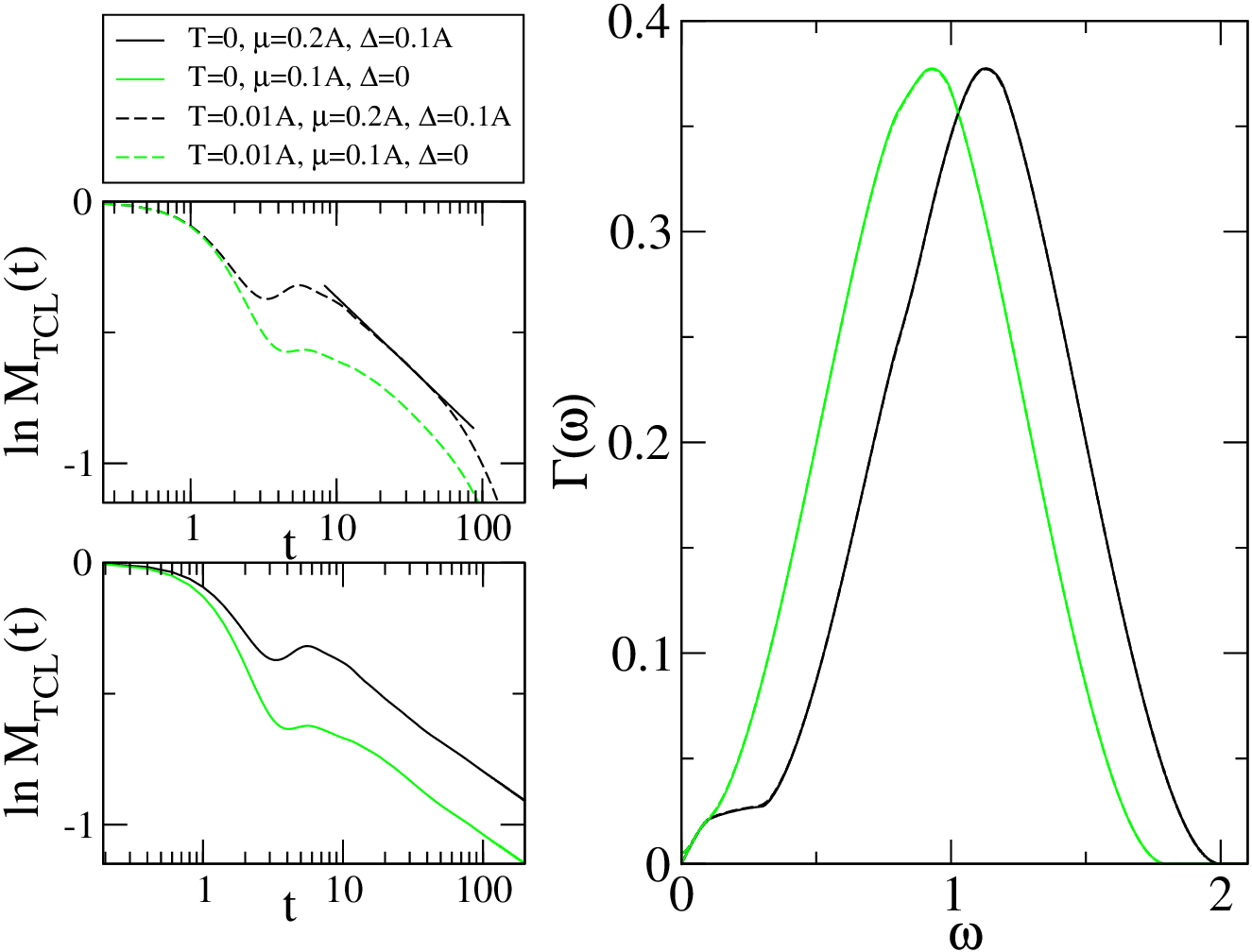}
%OLD\caption{Left panel: $\ln M_{TLC}/8\pi\lambda^2N^2$ as a function of $t$ for $\D=0$ and $\D=0.1A$ at $T=0$ (bold-lines) and $T=0.01A$ (dashed-lines). Right panel: Full $\Gw/8\pi\lambda^2N^2$ as a function of $\omega$ for $\D=0$ and $\D=0.1A$ at $T=0$.}
\caption{Left panel: $\ln M_{TLC}/8\pi\lambda^2N^2$ as a function of $t$ for $\D=0$ (green) and $\D=0.1A$ (black) at $T=0$ (bold lines) and $T=0.01A$ (dashed lines). Right panel: Full $\Gw/8\pi\lambda^2N^2$ as a function of $\omega$ for $\D=0$ and $\D=0.1A$ at $T=0$.}
\label{figura2_v7}
\end{figure}

 At finite temperature, we see two principal features:  
$\G(\w=0) \neq 0$  and $\G(\w)$ is fully analytic.  These features are also seen for  the square root DOS but one encounters non-analyticities  for the uniform DOS because of the discontinuities in the band structure. However, $\G$ 
is expected to be analytic for realistic metallic DOS.
 For a given band filling, the low frequency behavior of $\Gamma$ is  found to be qualitatively the same for both
 square root and elliptical density of states, implying that the asymptotic behavior is  indeed similar in both
 cases. For $T\neq 0$,    both  approximations TCL and NZ predict a 
Markovian decay of the coherence for times $t \gg t_T\sim1/T$, with a decay rate
 $\gamma=\G(0)$  as shown in Fig.~\ref{lnm}.    The equivalence of TCL and NZ at
 all times is clearly illustrated in the inset of  Fig. \ref{lnm}.   For a metallic DOS, the  decoherence rate 
 $\bar {\gamma}
 \propto  T$
  at low T and 
 $\bar {\gamma} \propto T\tanh A/T$ as $T\to \infty$ where $A$ is the cut-off. For $\rho_{un}(E)$ at low temperatures the exact result is  ${\bar{\gamma}}=2T(2-e^{-(\mu-\D)/T})$. Note that the
 temperature independent 
 proportionality factors are dependent on the details of the bath DOS.
 In  Fig. \ref{figura4_v7}, we plot the rate $\bar {\gamma}$   for metallic bath $\mu>\D$ for both square root and elliptical DOS.  
 As seen in the case of interacting spin baths \cite{Camalet2007a},  $\bar{ \gamma}$ increases with temperature and saturates  to a  finite DOS-dependent value proportional to  $\int{\rho(E)^2dE}$ at high temperatures.  This  saturation stems from the fact that the electrons have a spin degree of
freedom and  is different from the case of a bosonic bath, where the decoherence
rate never saturates. 
Fig.\ref{lnm} also shows that a higher chemical potential results in a faster decoherence at all  temperatures as there are more carriers which can dissipate energy. 
To summarize, in the weak coupling limit, both second order  TCL and NZ schemes predict that  a  simple metallic bath results in a power law decay of the asymptotic coherence at $T=0$ and a
Markovian decay at $T\neq 0$ irrespective of the nature of the underlying DOS. The corresponding exponents
and the Markovian rate do not depend very much on the qualitative details of the DOS. However, we expect the intermediate and other short time regimes to be indeed dependent on the details
of the DOS. \\
\noindent
To understand whether the presence of a gap in the DOS below the Fermi energy has any impact whatsoever on the decoherence,
we plot  $\G$  in  the right panel of Fig. \ref{figura2_v7} for the elliptic DOS  for $\D=0$ and $\D\neq0$ at the
 same value of the filling. At $T=0$, though the low frequency behavior  is the same in both cases,  a plateau like structure is indeed seen  in the vicinity of
 $\omega \sim 2 \Delta$ when $\D \neq 0$.   We observe  that  for asymptotic times,    $ M(t)  \propto  C t^{-\kappa}$  where $C$ is some constant which depends on the bath parameters and $\lambda$. We find that though the exponent $\kappa$ is the same irrespective of the value of $\D$, the coefficient $C$ is indeed dependent on $\D$ resulting in a faster decoherence in the gapless case cf. top left panel of  Fig. \ref{figura2_v7}.  At  low temperatures,  as seen in the bottom left panel of  Fig. \ref{figura2_v7}, though both DOS have an asymptotic Markovian regime with the same decay rate, the DOS with a gap has an
 intermediate power law regime for times $t_\D \sim 1/\D <  t \ll t_T$  which is not present for the  gapless metallic bath because $t_\D\sim \inn$.   This intermediate regime can be quite large depending on the parameters of the problem. To summarize, a gap in the DOS below the Fermi energy does serve  to reduce decoherence as compared to the case of a gapless DOS and this feature could probably be  exploited in  experiments. \\

\subsection{Semiconductor bath}
\label{sec:semi}
 We now focus on the semiconductor bath with a gap at the fermi level.  At zero temperature, $\Gw=0$ for $\omega \leq 2\D$ and $\G(\w) \propto\lambda^2 (\w-2\D)^2$  for $\omega$ close to $2\Delta$  for  all  DOS.
 The full behavior of $\Gw$ for the elliptical DOS  is plotted  in Fig. \ref{g_figure}.  We
 first focus on the TCL results.
At $T=0$,  we find that due to the presence of a gap in $\G$, the central spin only decoheres partially.
A straightforward evaluation of (\ref{eq:M_TCL})   shows the existence of three regimes:  a short time regime  $t_A \ll 1/A$ where the  central spin initially decoheres as
 a Gaussian 
 followed  by an oscillatory regime for $t < t_\D$  with a well
 defined minimum around $\overline{t}\simeq4/(2\D+A)$ and the asymptotic regime for   $t> t_\D$, where
 $M(t) \to const$ as a power law. These three regimes are expected to be generic and are indeed
 seen in the full numerical results for the coherence for both square root and
 elliptical DOS (see
 Fig. \ref{lnm}).

As temperature increases, the thermal activation of the gap results in a $\G$ which is no longer gapped  (cf. Fig. \ref{g_figure}).  Within the TCL formalism, this immediately implies that the asymptotic  behavior is Markovian $\ln{M_{TCL}(t)}\simeq-\gamma t$. 
  For low enough temperatures $T\ll\D$ which is the real semiconducting regime, we have the usual Gaussian behavior for $t < t_A$, followed by an oscillatory regime for $t_A \ll t \ll t_\D$, a power-law intermediate regime for $t_\D \ll t \ll t_T$ and finally the asymptotic  Markovian  regime for $t>t_T$ (cf. Fig. \ref{lnm}).  At high temperatures $T \gg \D$, there is no difference between the metal and the semiconductor and the intermediate  power law regime ceases to exist.

In contrast to the metallic case,  where the decoherence rate seemed to have 
the same qualitative  behavior for all the DOS considered,  the rate in the semiconducting case seems to be DOS dependent as $T \to 0$.  In this limit,  $\bar {\gamma}\propto T^2\exp{-(\D-\mu)/T}$ for $\rho_{sq}(E)$ and $\rho_{el}(E)$
 and $\bar{\gamma} \propto T\exp{-(\D-\mu)/T}$ for $\rho_{un}(E)$.  
Comparing with the case of the metallic bath where $\bar {\gamma} \propto T$, we see that the
 semiconducting bath  has a much smaller rate and hence longer coherence times which makes it a
 better environment for qubits than a simple metal. 
 However, as $T$ increases, $\bar{\gamma}$ has the same qualitative behavior as indicated in Fig. \ref{figura4_v7}.
 At higher temperatures, since the gap is completely smeared by thermal effects as expected  there is
 no real difference between the metallic and the semiconducting baths.
 
We now address the question of whether NZ predicts similar results for the asymptotic decoherence in the
semiconducting case.   Substituting  $\Lambda(\w)$ obtained from Eq. (\ref{lambdita})  in (\ref{gtilde}), we find that $\Gt(\w)$  typically has a three peak structure. 
  At low temperatures, $\Gt(\w)$ can be rewritten as the sum of two disjoint contributions $\Gt(\w)=\Gt^{l}(\w)+\Gt^{s}(\w)$  where $\Gt^{l}$ is a  Lorentzian peak  at low frequencies  and $\Gt^{s}$ describes  two  small satellite peaks which exist for $\omega \ge 2\Delta$. This separation of spectral weight in $\Gt$ 
leads to   $M_{NZ}(t)=M_{NZ}^{l}(t)+M_{NZ}^{s}(t)$ where  the first term is the usual Markovian decay $M_{NZ}^{l}(t) \simeq \exp{-\gamma t}$  with $\gamma$  being the same as that given by the TCL approximation.  To estimate the correction $M_{NZ}^{s}(t)$ stemming from the satellite  peaks, we note that  for $\w\gtrsim2\D$, $\Lambda(\varepsilon)\simeq\Lambda_0+\varepsilon\Lambda_1+\varepsilon^2\Lambda_2+\varepsilon^2\ln\varepsilon+O(\varepsilon^3)$ where $\varepsilon= \omega -2 \Delta$. 
Consequently, 
\begin{equation}\label{Greg}
\Gt^{s}(\w)\simeq\frac{\pi^2\lambda^2}{(\LL_0-2\D)^2}(\w-2\D)^2
\end{equation}
Using (\ref{Greg}) and (\ref{eq:M_NZ}), we find  that for long times 
\begin{align}\label{mnzsemi}
%M_{NZ}^{reg}(t)/\pi^2\lambda^2=&(\w-2\D)^2\sin{\w t}/t+\\&{2}(\w-2\D)\cos{\w t}/t^2-{2}\sin{\w t}/{t^3}\nonumber
M_{NZ}^{s}(t)/\pi\lambda^2 \sim \frac{1}{4\D^2}&(E_c-2\D)^2\sin{(E_c t)}/t\end{align}
where $E_c\gtrsim 2\D$ is a  cut-off scale for which $\Gt^{s}(\w)$ has a quadratic behavior. 
The true asymptotic behavior predicted by the NZ scheme  now depends on the competition between the
 exponential and the power law terms.  This defines a new time scale $t_{nz}\propto 1/\gamma \log (1/B\gamma)$ where 
$B$ is the pre-factor of the power law term seen in Eq.(\ref{mnzsemi}) such that for
$ t_T< t < t_{nz}$ one obtains the usual Markovian behavior, whereas for $t \gg t_{nz}$ one sees a non- Markovian power
law decay of the coherence.  In the weak coupling
formalism used here, $\lambda$ is expected to be much smaller than all the other scales of the hamiltonian, so a   $t_{nz}$ of the order of $\lambda^{-2}$ does correspond to very  very long times   for which the coherence has  already  decayed
to infinitesimal values and is practically unobservable. 
A full numerical integration of (\ref{eq:M_NZ}) with the appropriate $\Gt$ does show the
existence of these two asymptotic regimes $M_{NZ}^{s}(t)$ and $M_{NZ}^{l}(t)$
at low temperatures.  For  higher temperatures, the  gap in $\Gt$ is completely smeared by thermal fluctuations and one recovers the pure Markovian decay. 
The preceding analysis shows that the two approximations TCL and NZ do not necessarily predict same results
in the weak coupling limit even for simple environments like the semiconductor  considered in this section.
This divergence between the TCL and NZ approximations has  been seen in other problems like spin baths where
exact solutions are known \cite{Breuer2004}.  This shows that  one should exercise caution when using these approximation
schemes to calculate coherences in the cases of baths with more complicated characteristics.

\label{sec:spin}
\noindent
\begin{figure}[ht]
\centering
\includegraphics[trim=0cm 0cm -2cm -2cm,width=8cm]{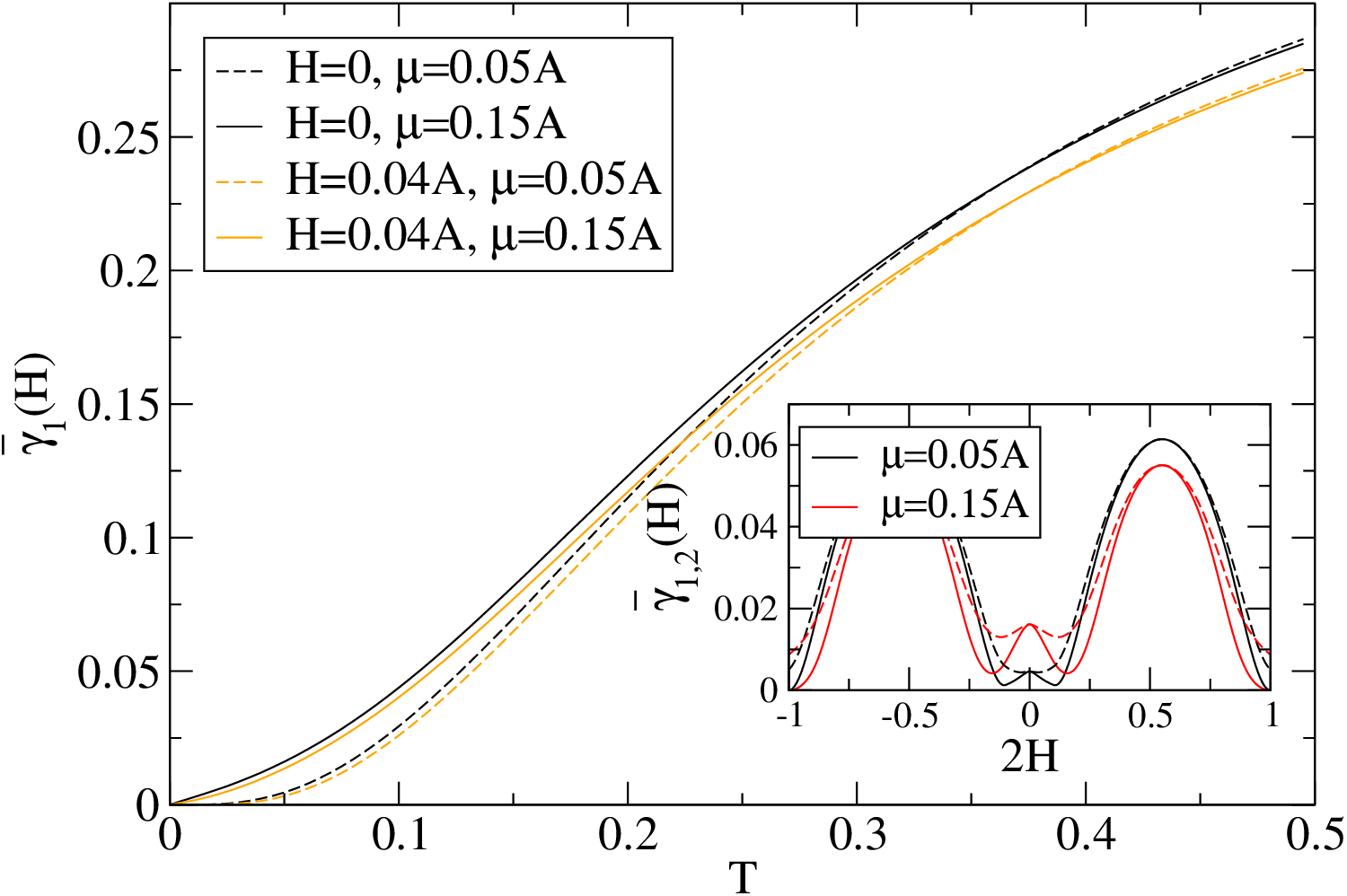}
%OLD\caption{(color online) Dimensionless relaxation rate $\bar{\gamma_1}(H)/N^2$ as a function of temperature in units of $A$ for a semiconductor $\mu=0.05A$ and a metal $\mu=0.15A$ for the elliptical DOS with $\D=0.1A$. Inset: Dimensionless rates $\bar{\gamma_{1,2}}(H)/N^2$ as a function of $2H$ for $T=0.05A$.}
\caption{(color online) Dimensionless relaxation rate $\bar{\gamma_1}(H)/N^2$ as a function of temperature in units of $A$ for a semiconductor $\mu=0.05A$ (dashed lines) and a metal $\mu=0.15A$ (bold lines) for fields $H=0$ (black) and $H=0.04A$ (orange) for the elliptical DOS with $\D=0.1A$. Inset: Dimensionless decoherence (dashed lines) and relaxation (bold lines) rates $\bar{\gamma_{1,2}}(H)/N^2$ as a function of $2H$ for a semiconductor  $\mu=0.05A$ (red) and a metal $\mu=0.15A$ (black) at $T=0.05A$.}
\label{figura5v7}
\end{figure}

\begin{figure}[ht]
\centering
\includegraphics[width=8cm]{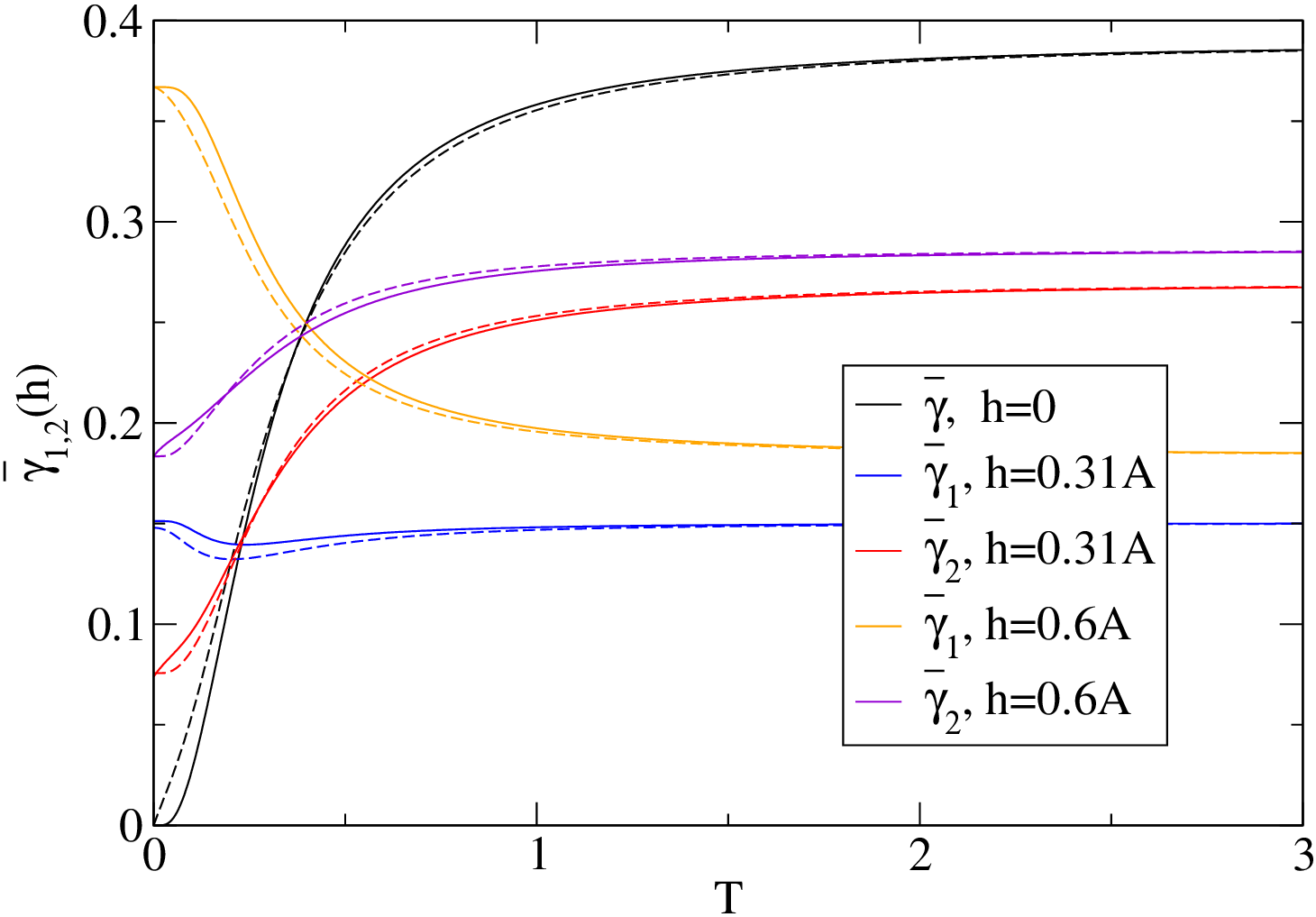}%{Figure.eps}%
\caption{Dimensionless decoherence and relaxation rates $\bar{\gamma}_{1,2}(h)/N^2$ for a semiconductor $\mu=0$ (dashed lines) and a metal $\mu=0.2A$ (bold lines) as a function of temperature for different values of $h$, $H=0$ and an elliptical DOS with $\D=0.1A$. Temperature is in units of $A$.}
\label{figura6_v7}
\end{figure}

\section{Non-zero magnetic field}
\label{sectionnon}
In this section,  we study the dependence of the asymptotic decoherence and relaxation rates on magnetic fields
   $H$ and $h$ applied  on the bath electrons and the central spin respectively cf. Eq.\ref{hamiltoniano}.  We explore whether an interplay between the magnetic fields  can be used as a way to control decoherence and relaxation.       Calculating the in-plane correlation functions $\langle \tilde{V}_{x,y}(t) \tilde{V}_{x,y}\rangle$  and the perpendicular correlation function   $\langle \tilde{V}_{z}(t) \tilde{V}_{z}\rangle$  and using them  in  Eqs. (\ref{eq:T1}) and (\ref{eq:T2}),   the dimensionless decoherence and relaxation rates are given by:
\begin{align}\label{rates}
{{\bar \gamma}_1}=\sum_{p=\pm1}&\int dE\rho(E)\rho(E+2p(H+h))\\
%&\times n(E+pH)(1-n(E+p(2h+H)))\nonumber\\
&\times n(E+pH)\left[1-n(E+p(2h+H))\right]\nonumber\\
{{\bar \gamma}_2}=\frac{1}{2}{{\bar \gamma}_1}+\frac{1}{2} \sum_{p=\pm1}&\int\rho(E)^2n(E+pH)\left[1-n(E+pH)\right]
\end{align} 
In the following, we assume  $h,H\geq0$.
We clearly see that for $H=h=0$, ${\bar{\gamma}}_1={{\bar{\gamma}}_2}$. For the special case of the Ising coupling ${\bf V}=(V_x,0,0)$ we have ${{\bar\gamma}_2}={{\bar \gamma}_1}/2$ for all  fields. 
  As we shall see below,  we have a rich spectrum of behavior  for the two rates as a function of field for all DOS.   Though the plots Fig.\ref{figura5v7}-\ref{figura7_v7}  show  the numerical results  for the elliptical density $\rho_{el}(E)$ only, we find that the key qualitative features summarized below are  true for all the DOS considered in this paper. The qualitative features of these results are seen for all the other DOS as well.  Additionally, we focus on  low and intermediate temperatures, since  the metallic ($\mu > \D$)  and semiconducting ($\mu <\D$)  baths are indistinguishable at high temperatures.
  %*********************************************
%*********************************************
\subsection{$h=0,H\neq 0$}

\begin{figure}[ht]  
\centering
\includegraphics[trim=0cm 0cm -2cm -2cm,width=8.5cm]{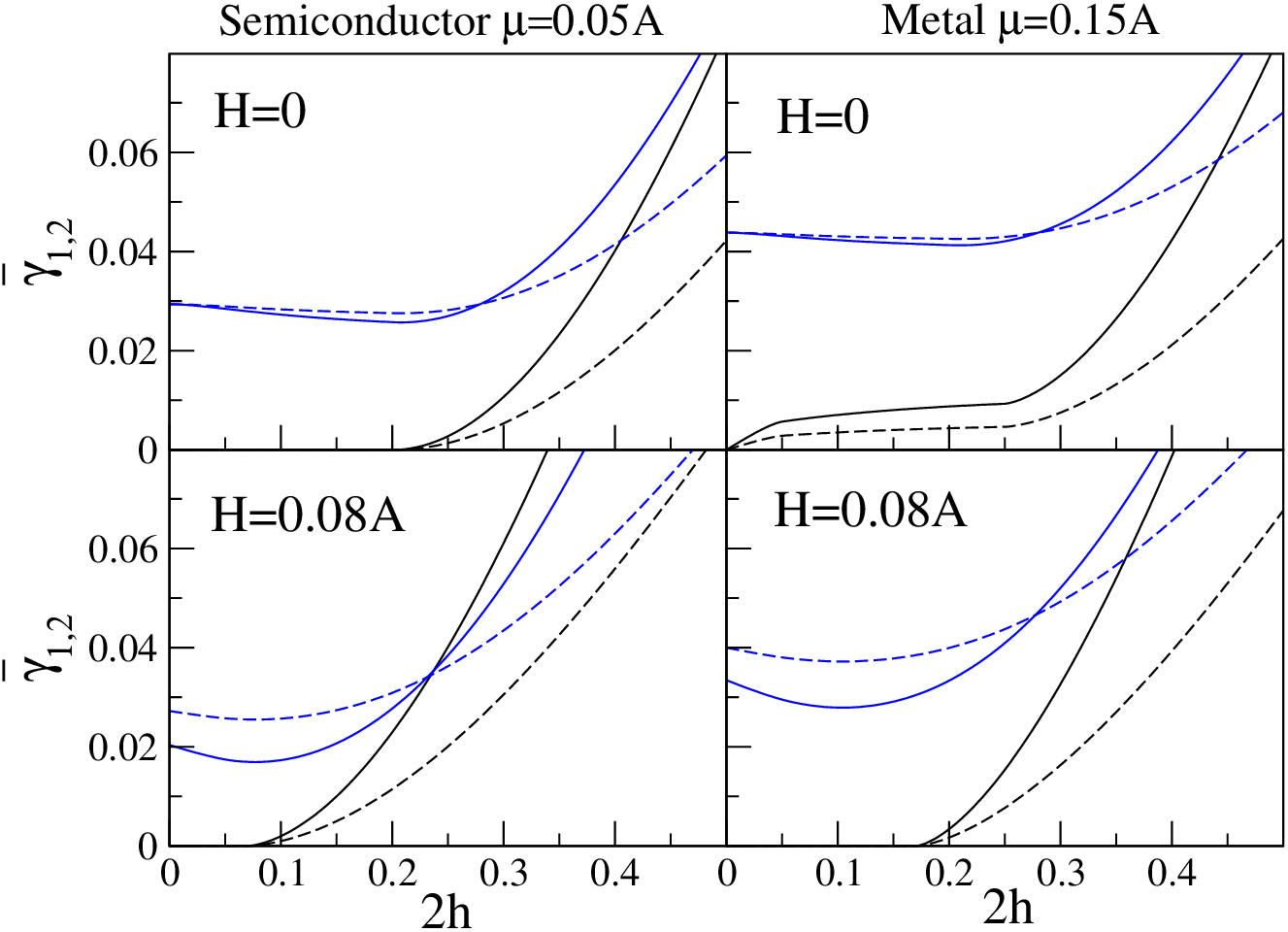}%{Figure.eps}%
\caption{(color online) Dimensionless relaxation (bold lines) and decoherence rate (dashed lines) $\bar{\gamma_{1,2}}/N^2$ as functions of $2h$ for a semiconductor bath $\mu=0.05A$ (left panel) and a metallic bath $\mu=0.15A$ (right panel) for $T=0$ (black) and $T=0.1A$ (blue) for the elliptical DOS with $\D=0.1A$ and $H=0$ (top panel) and $H=0.08A$ (bottom panel). }
\label{figura7_v7}
\end{figure}

 Our results for the two rates
$\bar{\gamma}_1$ and $\bar{\gamma}_2$ as a function of $H$ and $T$ are plotted in Figs. \ref{figura5v7}-\ref{figura7_v7} for both metallic and semiconducting baths. 
 We note that contrary to naive expectations, an external magnetic field applied on the bath does not always reduce the rates as compared to the $H=0$ case  and  in fact show a lot of anomalous behaviors.   As $T \to 0$,  both $\gb_1,\gb_2 \to 0$   for all $H$ indicating non-Markovian behavior. 
At finite  temperatures, as shown in the inset of  Fig. \ref{figura5v7},  
a common feature for the metallic and semiconducting baths is that both rates $\gb_1(H)$ and $\gb_2(H)$  have a non-monotonic dependence on  $H$ regardless of the value of the chemical potential $\mu$. This non-monotonicity of the rates stems from the presence of a gap in the DOS since at  high fields one is sensitive to the detailed nature of the DOS cf. Eq.(\ref{rates}).
 For example, such features are not seen in the case of a banal metallic bath described by  $\rho(E)\propto\Theta(A-|E|)$.

  At low temperatures, a straightforward calculation  shows that for the semiconductor,  ${\gb_{1}}^{s}(H) <{\gb}^{s}$ and for the metal, ${\gb_{1}}^{m}(H) <{\gb}^{m}$ if $H<\D$.  Here ${\gb}^{s}$ and ${\gb}^{m}$ denote the corresponding  $H=0$ rates for the semiconducting and metallic baths discussed in the previous sections.
 On the other hand though $\gb_2^s(H)<\gb^s$  and $\gb_2^m(H)<\gb^m$ 
   for $0<H<|\D-\mu|$, if $|\D-\mu|<H<\D $ then ${\gb_2}^{s}(H)>{\gb}^{s}$ and ${\gb_2}^{m}(H)<{\gb}^{m}$. (see Table \ref{tabla1}). This difference between the metallic and semiconducting baths  stems from  the fact that for $H>|\D-\mu|$,  changing the field is similar to changing the chemical potential (see eq. \ref{rates}) and the  bath switches from a metal-like behavior at low temperatures to a semiconductor-like behavior or viceversa. 
For extremely large fields $H\geq A$, the relaxation rate goes to zero as anticipated but the decoherence rate $\gb_2$  remains finite. This is true also for $h\neq0$ but the critical field for which it happens is $H=A-h$. 
  Note that in this high field limit  an Ising like coupling would lead to both
 rates $\gb_{1,2}=0$.  The Markovian decoherence  accompanied by zero rate of relaxation is a 
 manifestation of the  non-Ising nature of the coupling between the central spin and the bath. 
 To summarize,   fields $H<\D $ tend to partially suppress  decoherence and relaxation  induced by metallic baths.  However, in the case of semiconducting baths, though fields $H<\D-\mu$ do partially suppress relaxation and  decoherence,  fields $|\D-\mu|<H<\D $ tend to  augment decoherence.

\subsection{$H=0, h\neq 0$}

In this case,  where the central spin has intrinsic dynamics,  Eq.(\ref{rates})  shows that  ${\gb_1}=\G(\omega=2h)$  and 
${\gb_2}=\G(2h)/2+\G(0)/2$ (cf. Fig. \ref{g_figure} ).  Consequently,  both $\gb_1(h)$ and $\gb_2(h)$ have the same functional dependence  on the central spin field $h$.
We illustrate some scenarios where the  intrinsic dynamics play an interesting role in Fig. \ref{figura6_v7}.  For a metallic bath, turning on a small field $0<2h<|\D-\mu|$ at $T=0$ has the dramatic effect of transforming the non-Markovian decay of the decoherence and relaxation seen for $h=0$ into a Markovian decay since $\gb^m_{1,2} \propto h$ for small $h$. At low temperatures (see Table \ref{tabla1}), for fields $0<2h<\D -\mu$,  $\gb_{1,2}^{m}(h)>\gb^{m}$ whereas $\gb_{1,2}^{s}(h)<\gb^{s}$  due to the activation of the gap in $\G$ cf Fig. \ref{g_figure}. Moreover, these rates increase monotonically with $T$ and saturate to a finite value  as $T\rightarrow\inn$.

The  intermediate field regime yields the rather surprising result that for both metallic and semiconducting baths $\gb_1$ initially decreases with temperature,  and then  increases to attain a constant temperature independent value (blue curve Fig. \ref{figura6_v7}). However, $
\gb_2$ augments considerably in the same range of temperatures considered (red curve Fig. \ref{figura6_v7}). 
Additionally, there  exists a special value of the intrinsic field $h_c$ for which the relaxation rate (blue curve Fig \ref{figura6_v7}) has a reduced temperature dependence.   This last  feature also survives in the presence of a bath field $H$. For higher values of $h$, $\gb_1$ monotonically reduces with temperature but $\gb_2$ always increases with temperature (yellow and purple curves Fig \ref{figura6_v7}).   For all values of $h$,  we have  $\gamma_1 >\gamma_2$  for $T \lesssim \D$ and $\gamma_2 > \gamma_1$ for higher temperatures.  To our knowledge,  the possibility of an almost temperature independent relaxation at intermediate fields and the scenario where the relaxation rate decreases  and the decoherence rate increases with temperature have not been discussed in the literature.
\begin{table}[h]
\centering
\begin{tabular}{|c|c|c|}\hline
 Interval/bath & Metallic &  Semiconducting \\ \hline
\footnotesize $0<H<|\mu-\D|,h=0$ & \footnotesize${\gb_{1,2}}^{m}(H) <{\gb_{1,2}}^{m}$ &\footnotesize ${\gb_{1,2}}^{s}(H) <{\gb_{1,2}}^{s}$\\ \hline
\footnotesize$|\mu-\D|<H<\D,h=0$ & \footnotesize${\gb_{1,2}}^{m}(H) <{\gb_{1,2}}^{m}$ & \footnotesize${\gb_{1}}^{s}(H) <{\gb_{1}}^{s}$, ${\gb_{2}}^{s}(H) >{\gb_{2}}^{s}$\\ \hline
\footnotesize $0<2h<|\mu-\Delta|,H=0$ & \footnotesize${\gb_{1,2}}^{m}(h)  > {\gb_{1,2}}^{m}$ & \footnotesize${\gb_{1,2}}^{s}(h) <{\gb_{1,2}}^{s}$\\ \hline
\end{tabular}
\caption{Comparison between decoherence and relaxation rates $\gb_{1,2}^{m,s}(H,h)$ and $\gb_{1,2}^{m,s}$ for low temperatures and only one applied field.}
\label{tabla1}
\end{table}

\begin{figure}[ht]  
\centering
\includegraphics[width=8.5cm]{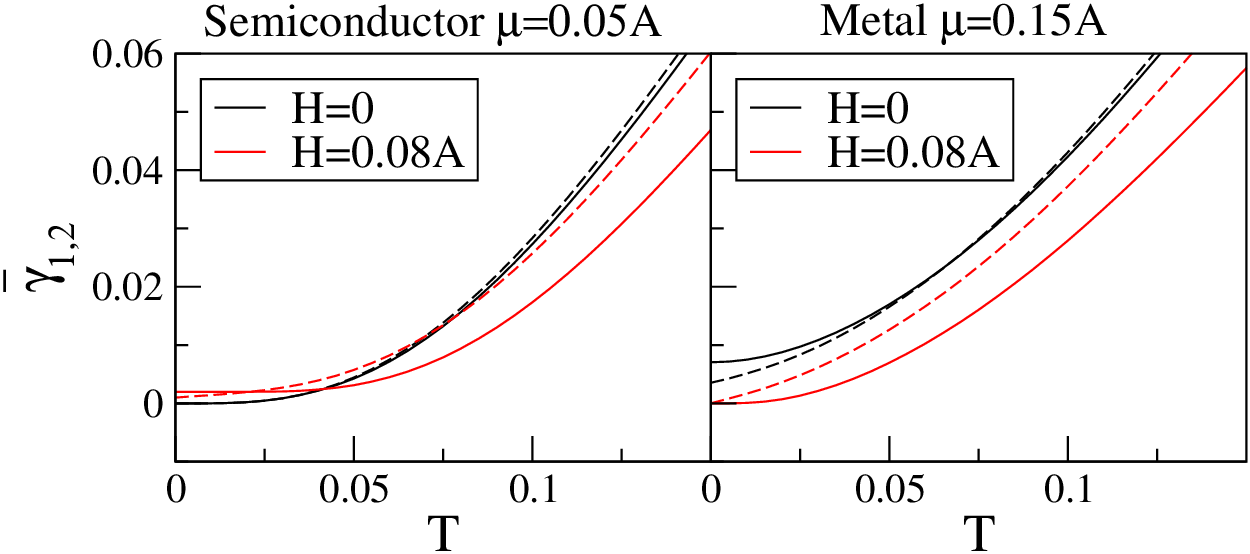}%{Figure.eps}%
\caption{(color online) Dimensionless relaxation (bold lines) and decoherence rate (dashed lines) $\bar{\gamma_{1,2}}/N^2$ as functions of $T$ for a semiconductor bath $\mu=0.05A$ (left panel) and a metallic bath $\mu=0.15A$ (right panel) for $H=0$ (black) and $H=0.08A$ (red) for the elliptical  DOS with $\D=0.1A$ and $2h=0.1A$. }
\label{figura8_v7}
\end{figure}   
%******************************************************************************************
%******************************************************************************************
\subsection{$h\neq 0,H\neq 0$}

In the presence of both fields, the low temperature behavior of the rates is an amalgam of the two cases studied above and is  sensitive to the nature of the bath.  The results for  the relaxation rate $\gb_1(H,h)$  are plotted in 
 Fig. \ref{figura7_v7} and  Fig. \ref{figura8_v7} for both baths.
  For a given value of $H$ and $T$, the decoherence  and relaxation rates  have the same functional dependence on $h$.   We now consider the $T=0$ case.  Firstly, we  note that at $T=0$, $\gb_1(H,h) >\gb_2(H,h)$ for all reasonable fields irrespective of the nature of the bath.  Moreover, for the metallic bath, if $0<H<|\D-\mu|$  then both $\gb_{1,2}$ are non-zero for any finite field $h$. If the field is further increased $|\D-\mu|<H<\D+\mu$, then $\gamma_{1,2}=0$  for
$2h<\mu+\D-H$. This is clearly illustrated in the  right panels of Fig. \ref{figura7_v7}.  
As a result,    the relaxation will  be Markovian or  non-Markovian  depending on  the values of $h$ and $H$. The case of the semiconducting bath is the opposite:  $\gb_{1,2}$  show a gap as a function of the  field $h$ for all  $H\leq \D+\mu$, indicating
non-Markovian behavior at low $h$ and  Markovian behavior for fields $2h > \D+\mu -H$  as shown in the left panels of Fig. \ref{figura7_v7}.  Finally, for  large  fields $H>\mu+\D$ any finite  $h$ leads to  finite  rates irrespective of the value of $\mu$.  
Consequently,   we have the surprising result that there exists a range of fields $max(\D+\mu_s -H,0)< 2h < \D+\mu_m -H$ and   $|\D-\mu_m|<H<\D+\mu_m$ where $\mu_s$ and $\mu_m$ denote the corresponding chemical potentials for the semiconducting and metallic baths,  for which  the  metallic bath is more effective   in suppressing relaxation and decoherence than the semiconducting bath. 
 This feature persists at low  temperatures $T \ll \D+\mu_s -H$, wherein there is a regime $h <h_1(T)$, where the semiconductor bath has lower rates  as expected 
 and a regime $h_1(T) < h < h_2(T)$, where  the metallic bath has lower rates as compared to the semiconductor. The interval $h_2 -h_1 \to 0$ as temperature increases.
  In the  first regime we find that $\gb_2 >\gb_1$ for all baths.
 We also see from Fig. \ref{figura8_v7} that for $0< H,h<|\D-\mu|$   ${\gb_1}^{s}(H,h)<{\gb_1}^{s}(H)<{\gb}^{s}$ and  $\gb_2^{s}(H,h)<\gb^{s}$. For larger fields $|\D-\mu|<H<\D$  and small enough $h$, $\gb_2^{s}(H,h)>\gb^{s}$ reminiscent of $h=0$ case. For a metallic bath, if $0<H<|\D-\mu|$  and $2h<\mu-\D-H$ then both the decoherence and the relaxation rate are greater than the decoherence and relaxation when no fields are present and we have $\gb_{1,2}^{m}(h)>\gb_{1,2}^{m}(H,h)>{\gb}^{m}$. 
    At   slightly higher temperatures, the semiconducting bath has a lower rate and at very high temperatures the two baths
are indistinguishable as expected.\\

 To summarize, we find that external magnetic fields do not necessarily
reduce the relaxation and decoherence rates of the qubit.  Depending on the nature of the bath and the values of the field, one sees a rich and varied behavior of the rates. For instance, in the absence of an intrinsic qubit field ($h=0$),
we find that the presence of a gap in the density of states leads to a non-monotonic variation of the rates as a function
of the bath field $H$.  In the presence of both fields,  $h$ and $H$,  an interplay  between the two magnetic fields leads
to a very interesting  regime  wherein for a qubit field  $h_1< h <h_2$,  the metallic bath has lower rates as compared to the semiconductor. The bounds $h_1$ and $h_2$ are functions of the gap $\D$, the bath field $H$ and the temperature.
For any given $H$, there exists a value of the bath field $h_c$ for which we have an almost temperature independent
relaxation.
 For very high bath fields, the  relaxation rate goes to zero but the decoherence rate remains finite signaling the non-Ising
 nature of the qubit-bath coupling.  It would be very interesting to see if these features survive in the presence of 
 an additional transverse field on the qubit.

\section{Discussion}  \label{sec:Diss}
We have studied the decoherence and relaxation of a central spin weakly coupled to a  bath of electrons described by different densities of states having a gap,  using two
well known approximation schemes, NZ and TCL. Though both methods predict an asymptotic Markovian decay for a metallic bath,
 we have  shown that  in the case of a semiconducting bath, the TCL  predicts a Markovian behavior at low $T$ whereas the NZ approximation predicts  non-Markovian behavior.  The  study of the validity  of both approximation techniques is left for future work. We emphasize that one has to exert care when using approximation schemes to  study the time evolution of the density matrix.   We find that a gap in the spectrum, even one away from the Fermi level,  generates an intermediate power law regime for the coherences at finite temperatures.  Depending on the parameters of the problem, this power law regime can be made sufficiently large so as to be relevant for experiments.   One might also question the validity of the weak coupling formalism used here since it is well known that even the
equilibrium physics of the Kondo model is governed  by  strong coupling physics, which leads to a complete
screening of the Kondo spin.
This  was 
 explored in Ref.\cite{Schiller2006}, where numerical renormalization group methods were used to study
 the decoherence of the Kondo spin. It was found that for any value of the coupling $\lambda$ there is a
 Kondo timescale $t_\lambda \propto \exp{const/\lambda}$ beyond which the Kondo spin gets screened
 dynamically.   The weak coupling results  discussed here were indeed valid for  {\it intermediate} times $t \ll t_\lambda$. In this paper, since we are only interested in  very small $\lambda$, we expect our 
 results to be valid for realistically long times since the time scale $t_\lambda \to \infty$ as $\lambda \to 0$. \\
   We have also calculated the relaxation and decoherence
  rate as a function of external magnetic fields applied on the bath and/or central spin.  We encounter
  novel situations where for moderately large fields on the central spin alone result in a  nearly temperature independent relaxation but a rapidly increasing decoherence. 
 We find that the presence of gap anywhere  in the density of states  has important ramifications for the rates.  This leads to the case where for a certain range of the bath field $H$, there is a regime  $h_1(T) < h < h_2(T)$, where we have the surprising result that  the metallic bath has lower rates of relaxation and decoherence as compared to the semiconductor.   To conclude,  there exists an interesting interplay between the gap in the density of states and the applied magnetic fields, which can have a lot of practical advantages for  experiments on qubits. 

 - 
  
\appendix
\section{ Self-energy}
\label{annex_self} 
Let us consider a general hamiltonian $H=H_S + H_B+  H_ {SB}$ where $H_S=-h\sigma_z$ and $H_{SB}=V_x\sigma_x+V_y\sigma_y+V_z\sigma_z$. The operators $V$ are proportional to $\lambda$ and $\lambda$ is smaller than any other scale of the system. The Kondo coupling considered in this article is a specific case.\\
By following a procedure similar to the one presented in the appendix of \cite{Camalet2007a} for the Ising coupling that uses the resolvent operator formalism \cite{Cohen} in the weak coupling regime one obtains the general form of the self energy matrix $\Sigma_{\beta\alpha} $ and $h_{\beta\alpha}$ in eq.  (\ref{eqM}). By definition $\Sigma_{0\alpha}(z)=0$ for $\alpha=x,y,z$ and $\Sigma_{00}(z)=1/z$ which assures the conservation of the trace. For the other components:
\begin{widetext}
\begin{eqnarray}
\Sigma_{\beta \alpha }(z) &=& -i \int_{0} ^{\infty} dt e^{izt} \sum_{\gamma,\delta} \left(1-\delta_{\beta\gamma} \right) 
\left\lbrace
\langle \tilde{V}_{\gamma}(t) \tilde{V}_{\delta} \rangle \mathrm{Tr} \left( \sigma_{c}^{\beta} \sigma_{c}^{\gamma} e^{ith\sigma_{c}^{z} } \sigma_{c}^{\delta} \sigma_{c}^{\alpha} e^{-ith\sigma_{c}^{z} }  \right) \right. \nonumber \\
& & + \left. \langle \tilde{V}_{\delta} \tilde{V}_{\gamma}(t) \rangle \mathrm{Tr} \left( \sigma_{c}^{\gamma} \sigma_{c}^{\beta} e^{ith\sigma_{c}^{z} } \sigma_{c}^{\alpha} \sigma_{c}^{\delta} e^{-ith\sigma_{c}^{z} }  \right)  \right\rbrace 
\label{self} 
\end{eqnarray}
\end{widetext}
The $h_{\beta\alpha}$ are given by:
\begin{equation}
h_{\beta\alpha}=\frac{1}{2}\mathrm{Tr}\left(\sigma_\beta \left[H_S+\langle H_{SB}\rangle,\sigma_\alpha\right]\right)
\end{equation}
Where $Tr$ it the total trace. For the specific case of the Kondo coupling:
\begin{equation}
h_{\beta\alpha}=\frac{1}{2}\mathrm{Tr}\left(\sigma_\beta \left[H_S,\sigma_\alpha\right]\right)
\end{equation}
since $\langle V_x\rangle=\langle V_y\rangle=\langle V_z\rangle=0$ implying $\langle H_{SB}\rangle=0$ for $H=0$.

%\bibliography{semi_final}

\begin{thebibliography}{23}
\expandafter\ifx\csname natexlab\endcsname\relax\def\natexlab#1{#1}\fi
\expandafter\ifx\csname bibnamefont\endcsname\relax
  \def\bibnamefont#1{#1}\fi
\expandafter\ifx\csname bibfnamefont\endcsname\relax
  \def\bibfnamefont#1{#1}\fi
\expandafter\ifx\csname citenamefont\endcsname\relax
  \def\citenamefont#1{#1}\fi
\expandafter\ifx\csname url\endcsname\relax
  \def\url#1{\texttt{#1}}\fi
\expandafter\ifx\csname urlprefix\endcsname\relax\def\urlprefix{URL }\fi
\providecommand{\bibinfo}[2]{#2}
\providecommand{\eprint}[2][]{\url{#2}}

\bibitem[{\citenamefont{{Nielsen} and {Chuang}}(2000)}]{Nielsen2000}
\bibinfo{author}{\bibfnamefont{M.~A.} \bibnamefont{{Nielsen}}}
  \bibnamefont{and} \bibinfo{author}{\bibfnamefont{I.~L.}
  \bibnamefont{{Chuang}}}, \emph{\bibinfo{title}{{Quantum Computation and
  Quantum Information.}}} (\bibinfo{publisher}{Cambridge University Press},
  \bibinfo{year}{2000}).

\bibitem[{\citenamefont{{Bose}}(2003)}]{Bose2003}
\bibinfo{author}{\bibfnamefont{S.}~\bibnamefont{{Bose}}},
  \bibinfo{journal}{Physical Review Letters} \textbf{\bibinfo{volume}{91}},
  \bibinfo{pages}{207901} (\bibinfo{year}{2003}).

\bibitem[{\citenamefont{Makhlin et~al.}(2001)\citenamefont{Makhlin, Sch\"on,
  and Shnirman}}]{Makhlin2001}
\bibinfo{author}{\bibfnamefont{Y.}~\bibnamefont{Makhlin}},
  \bibinfo{author}{\bibfnamefont{G.}~\bibnamefont{Sch\"on}}, \bibnamefont{and}
  \bibinfo{author}{\bibfnamefont{A.}~\bibnamefont{Shnirman}},
  \bibinfo{journal}{Rev. Mod. Phys.} \textbf{\bibinfo{volume}{73}},
  \bibinfo{pages}{357} (\bibinfo{year}{2001}).

\bibitem[{\citenamefont{Petta et~al.}(2004)\citenamefont{Petta, Johnson,
  Marcus, Hanson, and Gossard}}]{Petta2004}
\bibinfo{author}{\bibfnamefont{J.~R.} \bibnamefont{Petta}},
  \bibinfo{author}{\bibfnamefont{A.~C.} \bibnamefont{Johnson}},
  \bibinfo{author}{\bibfnamefont{C.~M.} \bibnamefont{Marcus}},
  \bibinfo{author}{\bibfnamefont{M.~P.} \bibnamefont{Hanson}},
  \bibnamefont{and} \bibinfo{author}{\bibfnamefont{A.~C.}
  \bibnamefont{Gossard}}, \bibinfo{journal}{Physical Review Letters}
  \textbf{\bibinfo{volume}{93}}, \bibinfo{eid}{186802}
  (pages~\bibinfo{numpages}{4}) (\bibinfo{year}{2004}).

\bibitem[{\citenamefont{Caldeira and Leggett}(1981)}]{Leggett1981}
\bibinfo{author}{\bibfnamefont{A.~O.} \bibnamefont{Caldeira}} \bibnamefont{and}
  \bibinfo{author}{\bibfnamefont{A.~J.} \bibnamefont{Leggett}},
  \bibinfo{journal}{Phys. Rev. Lett.} \textbf{\bibinfo{volume}{46}},
  \bibinfo{pages}{211} (\bibinfo{year}{1981}).

\bibitem[{\citenamefont{{Leggett} et~al.}(1987)\citenamefont{{Leggett},
  {Chakravarty}, {Dorsey}, {Fisher}, {Garg}, and {Zwerger}}}]{Leggett1987}
\bibinfo{author}{\bibfnamefont{A.~J.} \bibnamefont{{Leggett}}},
  \bibinfo{author}{\bibfnamefont{S.}~\bibnamefont{{Chakravarty}}},
  \bibinfo{author}{\bibfnamefont{A.~T.} \bibnamefont{{Dorsey}}},
  \bibinfo{author}{\bibfnamefont{M.~P.~A.} \bibnamefont{{Fisher}}},
  \bibinfo{author}{\bibfnamefont{A.}~\bibnamefont{{Garg}}}, \bibnamefont{and}
  \bibinfo{author}{\bibfnamefont{W.}~\bibnamefont{{Zwerger}}},
  \bibinfo{journal}{Rev. Mod. Phys.} \textbf{\bibinfo{volume}{59}},
  \bibinfo{pages}{1} (\bibinfo{year}{1987}).

\bibitem[{\citenamefont{Camalet and Chitra}(2007{\natexlab{a}})}]{Camalet2007}
\bibinfo{author}{\bibfnamefont{S.}~\bibnamefont{Camalet}} \bibnamefont{and}
  \bibinfo{author}{\bibfnamefont{R.}~\bibnamefont{Chitra}},
  \bibinfo{journal}{Phys. Rev. Lett.} \textbf{\bibinfo{volume}{99}},
  \bibinfo{pages}{267202} (\bibinfo{year}{2007}{\natexlab{a}}).

\bibitem[{\citenamefont{Camalet and Chitra}(2007{\natexlab{b}})}]{Camalet2007a}
\bibinfo{author}{\bibfnamefont{S.}~\bibnamefont{Camalet}} \bibnamefont{and}
  \bibinfo{author}{\bibfnamefont{R.}~\bibnamefont{Chitra}},
  \bibinfo{journal}{Phys. Rev. B} \textbf{\bibinfo{volume}{75}},
  \bibinfo{pages}{094434} (\bibinfo{year}{2007}{\natexlab{b}}).

\bibitem[{\citenamefont{Anders and Schiller}(2006)}]{Schiller2006}
\bibinfo{author}{\bibfnamefont{F.~B.} \bibnamefont{Anders}} \bibnamefont{and}
  \bibinfo{author}{\bibfnamefont{A.}~\bibnamefont{Schiller}},
  \bibinfo{journal}{Phys. Rev. B} \textbf{\bibinfo{volume}{74}},
  \bibinfo{pages}{245113} (\bibinfo{year}{2006}).

\bibitem[{\citenamefont{Stamp and Tupitsyn}(2004)}]{Stamp2004}
\bibinfo{author}{\bibfnamefont{P.~C.~E.} \bibnamefont{Stamp}} \bibnamefont{and}
  \bibinfo{author}{\bibfnamefont{I.~S.} \bibnamefont{Tupitsyn}},
  \bibinfo{journal}{Phys. Rev. B} \textbf{\bibinfo{volume}{69}},
  \bibinfo{pages}{014401} (\bibinfo{year}{2004}).

\bibitem[{\citenamefont{Rossini et~al.}(2007)\citenamefont{Rossini, Calarco,
  Giovannetti, Montangero, and Fazio}}]{Rossini2007}
\bibinfo{author}{\bibfnamefont{D.}~\bibnamefont{Rossini}},
  \bibinfo{author}{\bibfnamefont{T.}~\bibnamefont{Calarco}},
  \bibinfo{author}{\bibfnamefont{V.}~\bibnamefont{Giovannetti}},
  \bibinfo{author}{\bibfnamefont{S.}~\bibnamefont{Montangero}},
  \bibnamefont{and} \bibinfo{author}{\bibfnamefont{R.}~\bibnamefont{Fazio}},
  \bibinfo{journal}{Phys. Rev. A} \textbf{\bibinfo{volume}{75}},
  \bibinfo{pages}{032333} (\bibinfo{year}{2007}).

\bibitem[{\citenamefont{Witzel and {Das Sarma}}(2006)}]{DasSarma2006}
\bibinfo{author}{\bibfnamefont{W.~M.} \bibnamefont{Witzel}} \bibnamefont{and}
  \bibinfo{author}{\bibfnamefont{S.}~\bibnamefont{{Das Sarma}}},
  \bibinfo{journal}{Phys. Rev. B} \textbf{\bibinfo{volume}{74}},
  \bibinfo{pages}{035322} (\bibinfo{year}{2006}).

\bibitem[{\citenamefont{Yamada et~al.}(2007)\citenamefont{Yamada, Sakuma, and
  Tsuchiura}}]{Yamada2007}
\bibinfo{author}{\bibfnamefont{N.}~\bibnamefont{Yamada}},
  \bibinfo{author}{\bibfnamefont{A.}~\bibnamefont{Sakuma}}, \bibnamefont{and}
  \bibinfo{author}{\bibfnamefont{H.}~\bibnamefont{Tsuchiura}},
  \bibinfo{journal}{J. Appl. Phys.} \textbf{\bibinfo{volume}{101}}
  (\bibinfo{year}{2007}), ISSN \bibinfo{issn}{{0021-8979}}.

\bibitem[{\citenamefont{Gao and Xiong}(2008)}]{Gao2008}
\bibinfo{author}{\bibfnamefont{Y.}~\bibnamefont{Gao}} \bibnamefont{and}
  \bibinfo{author}{\bibfnamefont{S.-J.} \bibnamefont{Xiong}},
  \bibinfo{journal}{Physics Letters A} \textbf{\bibinfo{volume}{372}},
  \bibinfo{pages}{4630} (\bibinfo{year}{2008}), ISSN \bibinfo{issn}{0375-9601}.

\bibitem[{\citenamefont{Lutchyn et~al.}(2008)\citenamefont{Lutchyn,
  Cywi\ifmmode~\acute{n}\else \'{n}\fi{}ski, Nave, and
  Das~Sarma}}]{Lutchyn2008}
\bibinfo{author}{\bibfnamefont{R.~M.} \bibnamefont{Lutchyn}},
  \bibinfo{author}{\bibfnamefont{L.}~\bibnamefont{Cywi\ifmmode~\acute{n}\else
  \'{n}\fi{}ski}}, \bibinfo{author}{\bibfnamefont{C.~P.} \bibnamefont{Nave}},
  \bibnamefont{and}
  \bibinfo{author}{\bibfnamefont{S.}~\bibnamefont{Das~Sarma}},
  \bibinfo{journal}{Phys. Rev. B} \textbf{\bibinfo{volume}{78}},
  \bibinfo{pages}{024508} (\bibinfo{year}{2008}).

\bibitem[{\citenamefont{Winograd et~al.}(2009)\citenamefont{Winograd,
  Rozenberg, and Chitra}}]{Winograd2009}
\bibinfo{author}{\bibfnamefont{E.~A.} \bibnamefont{Winograd}},
  \bibinfo{author}{\bibfnamefont{M.~J.} \bibnamefont{Rozenberg}},
  \bibnamefont{and} \bibinfo{author}{\bibfnamefont{R.}~\bibnamefont{Chitra}},
  \bibinfo{journal}{Phys. Rev. B} \textbf{\bibinfo{volume}{80}},
  \bibinfo{pages}{214429} (\bibinfo{year}{2009}).

\bibitem[{\citenamefont{Breuer and Petruccione}(2002)}]{Breuer}
\bibinfo{author}{\bibfnamefont{H.~P.} \bibnamefont{Breuer}} \bibnamefont{and}
  \bibinfo{author}{\bibfnamefont{F.}~\bibnamefont{Petruccione}},
  \emph{\bibinfo{title}{The theory of Open Quantum Systems}}
  (\bibinfo{publisher}{Oxford University Press}, \bibinfo{year}{2002}).

\bibitem[{\citenamefont{Weiss}(2008)}]{Weiss}
\bibinfo{author}{\bibfnamefont{U.}~\bibnamefont{Weiss}},
  \emph{\bibinfo{title}{Quantum dissipative systems}}, Series in modern
  condensed matter physics (\bibinfo{publisher}{World Scientific},
  \bibinfo{year}{2008}), ISBN \bibinfo{isbn}{9789812791627}.

\bibitem[{\citenamefont{{Dattagupta} et~al.}(1989)\citenamefont{{Dattagupta},
  {Grabert}, and {Jung}}}]{Dattagupta1989}
\bibinfo{author}{\bibfnamefont{S.}~\bibnamefont{{Dattagupta}}},
  \bibinfo{author}{\bibfnamefont{H.}~\bibnamefont{{Grabert}}},
  \bibnamefont{and} \bibinfo{author}{\bibfnamefont{R.}~\bibnamefont{{Jung}}},
  \bibinfo{journal}{Journal of Physics Condensed Matter}
  \textbf{\bibinfo{volume}{1}}, \bibinfo{pages}{1405} (\bibinfo{year}{1989}).

\bibitem[{\citenamefont{Breuer et~al.}(2004)\citenamefont{Breuer, Burgarth, and
  Petruccione}}]{Breuer2004}
\bibinfo{author}{\bibfnamefont{H.-P.} \bibnamefont{Breuer}},
  \bibinfo{author}{\bibfnamefont{D.}~\bibnamefont{Burgarth}}, \bibnamefont{and}
  \bibinfo{author}{\bibfnamefont{F.}~\bibnamefont{Petruccione}},
  \bibinfo{journal}{Phys. Rev. B} \textbf{\bibinfo{volume}{70}},
  \bibinfo{pages}{045323} (\bibinfo{year}{2004}).

\bibitem[{\citenamefont{Abragam}(1994)}]{Abragam1994}
\bibinfo{author}{\bibfnamefont{A.}~\bibnamefont{Abragam}},
  \emph{\bibinfo{title}{The principles of nuclear magnetism}}, International
  Series of Monographs on Physics (\bibinfo{publisher}{Oxford University
  Press}, \bibinfo{year}{1994}), ISBN \bibinfo{isbn}{9780198520146}.

\bibitem[{\citenamefont{Kittel}(1971)}]{Kittel}
\bibinfo{author}{\bibfnamefont{C.}~\bibnamefont{Kittel}},
  \emph{\bibinfo{title}{Introduction to Solid State Physics}}
  (\bibinfo{publisher}{Wiley}, \bibinfo{address}{New York},
  \bibinfo{year}{1971}), \bibinfo{edition}{4th} ed., ISBN
  \bibinfo{isbn}{0-471-49021-0}.

\bibitem[{\citenamefont{Claude Cohen-Tannoudji}(1998)}]{Cohen}
\bibinfo{author}{\bibfnamefont{G.~G.} \bibnamefont{Claude Cohen-Tannoudji},
  \bibfnamefont{Jacques Dupont-Roc}}, \emph{\bibinfo{title}{Atom-Photon
  Interactions: Basic Processes and Applications}} (\bibinfo{year}{1998}).

\end{thebibliography}

\end{document}